\input{aipcheck}

\documentclass[
    ,final            
  ]
  {aipproc}

\layoutstyle{6x9}

\def \src {\mbox{IGR\,J16418--4532}}
\def \inte {\emph{INTEGRAL}}
\def \xmm {\emph{XMM--Newton}}

\def \sw {{\it Swift}}
\def \rxte {\emph{RXTE}}

\def \hcm {\hbox {\ifmmode $ atom cm$^{-2}\else atom cm$^{-2}$\fi}}

\def \ATel {Astron.~Tel.}
\def \apj {ApJ}

\def \apjs {ApJS}
\def \aj {AJ}
\def \aap {A\&A}

\def \mnras {MNRAS}

\usepackage[font=small,labelfont=bf]{caption}
\begin{document}

\title{\emph{Swift} monitoring of IGR~J16418--4532}


\author{P.~Esposito}{address={INAF -- Istituto di Astrofisica Spaziale e Fisica Cosmica - Milano, Italy}}

\author{P.~Romano}{address={INAF -- Istituto di Astrofisica Spaziale e Fisica Cosmica - Palermo, Italy}}

\author{V. Mangano}{address={INAF -- Istituto di Astrofisica Spaziale e Fisica Cosmica - Palermo, Italy}}

\author{L. Ducci}{address={Institut f\"ur Astronomie und Astrophysik, Universit\"at T\"ubingen, Germany}}

\author{S. Vercellone}{address={INAF -- Istituto di Astrofisica Spaziale e Fisica Cosmica - Palermo, Italy}}

\begin{abstract}
We report on the {\it Swift} observations of the candidate supergiant fast X-ray transient (SFXT) IGR~J16418--4532, which has an orbital  period of $\sim$3.7 d. Our monitoring, for a total of $\sim$43 ks, spans over three orbits and represents the most intense and complete sampling along the orbital period of the light curve of this source. If one assumes a circular orbit, the X-ray emission from this source can be explained by accretion from a spherically symmetric clumpy wind from a blue supergiant, composed of clumps with different masses, ranging from $\sim$$5\times10^{16}$~g to $10^{21}$~g. 
\end{abstract}

\maketitle


\section{Introduction}
  
Supergiant Fast X-ray Transients (SFXTs) are a class of high-mass X-ray binaries with OB supergiant companions. Their hallmark is the occurrence of short ($\sim$hours) outbursts during which the luminosity can increase by 3--5 orders of magnitude (typically, up to $10^{37}$ erg s$^{-1}$). Either the clumpy structure of the wind from the companion or a centrifugal and/or magnetic gating have been suggested to be responsible for the outbursts.

The hard-X-ray transient \src\ was discovered by \inte\ in 2003 and later tentatively classified as an SFXT on the basis of its outbursting behaviour \citep{Sguera2006}. At soft X-rays \xmm\ observations showed a heavily absorbed pulsar with $P_{\rm spin}\simeq1.25$ ks \citep{Walter2006}. An orbital periodicity of $\sim$3.74 d was discovered with the \sw/BAT and the \rxte/ASM \citep{Corbet2006:atel779,Levine2011} and a hint of a total eclipse,  consistent with a supergiant companion, has been observed in the BAT data. We analyzed all the \sw\ data collected on \src, including a monitoring campaign carried out in 2011 July that spans over three orbital periods (see \citep{romano12} for a complete log of the observations). 

 	 \section{Results \label{igr16418:results} }

\begin{figure}
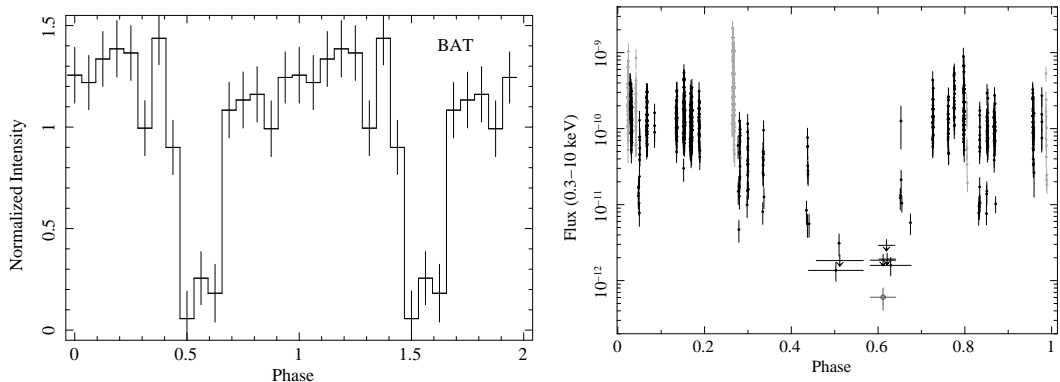

\includegraphics[height=7cm,angle=-90]{esposito_16418_fig1.ps}\includegraphics[height=7cm,angle=-90]{esposito_16418_fig2.ps}
\caption{\emph{Left panel:} BAT light curve folded at $P_{\rm orb}=3.73886$\,d and  $T_{\rm epoch}=$ MJD 53560.20000. \emph{Right Panel:} XRT 0.3--10\,keV flux light curve of \src, folded at the same period and epoch. The data were collected in 2011 February and June (grey), and in 2011 July (black). Downward-pointing arrows are 3$\sigma$ upper limits. The open circle is a detection obtained by combining three observations; it is the lowest point of the monitoring and corresponds to an unabsorbed 0.3--10\,keV flux of $\sim$$6.0\times10^{-13}$ erg cm$^{-2}$ s$^{-1}$. The peak count rate is reached on MJD 55610.08, at $2.1\times10^{-9}$ erg cm$^{-2}$ s$^{-1}$.\label{igr16418:fig:batlcvphased}}
\end{figure}

For the binary orbital period analysis we retrieved from the BAT Transient Monitor page \citep{Krimm2006_atel_BTM} the 15--50\,keV BAT light curves (averaged over \sw's orbital period of $\sim$5400\,s)  covering the data range from 2005 February 12 to 2011 July 12. Fig.\,\ref{igr16418:fig:batlcvphased}\,(left) shows the epoch-folded BAT light curve for $P_{\rm orb}=3.73886(28)$\,d and MJD 53560.20000 as $T_{\rm epoch}$ \citep{Corbet2006:atel779,Levine2011}. The BAT data show an eclipse centred at $\phi \sim 0.55$.  The depth is consistent with a total eclipse, and the eclipse FWHM duration is $\Delta\phi=0.17\pm0.05$ of the orbital period. 

Fig.\,\ref{igr16418:fig:batlcvphased}\,(right) shows the 0.3--10\,keV XRT epoch-folded and background-subtracted light curve of the whole 2011 July campaign. In order to convert count rate to flux, we adopted conversion factors derived from spectral fits of each observation (adopting a simple power-law model) and subsequent interpolation across observations where fitting was not feasible (see \citep{romano12} for more details). The data start at phase 0.16 and cover three full periods. Superimposed on the long-term orbital modulation,  which follows that seen with BAT, flaring is observed on short time scales (a behaviour typical of most SFXTs). In particular, the eclipse lasts $\Delta \phi \sim 0.2$ and is centred at $\phi \sim 0.55$. 

 	 \section{Discussion \label{igr16418:discussion}}

The \sw\ folded light curves of the source show a dip consistent with zero intensity. If interpreted as an eclipse, it can be used to infer the nature of the stellar companion. Excluding the eclipse, the BAT lightcurve does not show substantial flux modulation along the orbit, suggesting that the eccentricity is not large. This is in agreement with the observation that in many X-ray binary systems with nearly circular orbits and short orbital periods ($\leq$\,4\,days), the eccentric orbit produced by the explosion of the supernova is circularized on timescale of $10^6$\,yr (consistent with the evolution timescale of an O star), due to viscous tidal interactions. Thus, for simplicity, we assume a circular orbit ($P_{\rm orb}= 3.73886$\,d). Let us adopt the typical values of stellar masses and radii \citep{Martins2005} for a primary of spectral type O8.5 and luminosity classes I, III, or V, and for the neutron star a mass of $M_{\rm NS}=1.4$\,M$_\odot$, and a radius of $R_{\rm NS}=10$\,km. By combining Kepler's third law with the relationship given by \citet{Rappaport1983} for the radius of the primary $R$, $(R/a)^2= \cos^2 i +\sin^2 i \, \sin^2 \theta_{\rm e},$ where $a$ is the semi-major axis of the orbit, $i$ is the inclination, and $\theta_{\rm e}$ is the eclipse semi-angle, we find that the observed $\Delta\phi=0.2$ is inconsistent with a luminosity class III or V primary. We conclude that indeed the primary is an O8.5I supergiant star (with mass, radius, luminosity, effective temperature given by \citet{Martins2005}, $M_{\rm OB}=31.5$\,M$_\odot$, $R_{\rm OB}=21.4$\,R$_\odot$, $\log L_{\rm OB}/L_\odot=5.65$, $T_{\rm eff} = 32274$\,K) and that the distance to \src\ is $\sim$13\,kpc \citep{Rahoui2008}. 

The XRT light curve shows a large dynamic range, up to 1--2 orders of magnitude within a single snapshot and of at least 3 orders of magnitude overall (even when the low-flow flux measured during the eclipse is excluded). While this dynamic range is lower than that observed in the classical SFXTs, which reach 4--5 orders of magnitude in dynamic range, it is indeed typical of intermediate SFXTs (e.g. IGR\,J18483$-$0311). The 2011 February flare (at phase $\sim$0.27 in Fig.\,\ref{igr16418:fig:batlcvphased}), in particular, reached $\sim 2\times10^{-9}$\,erg cm$^{-2}$ s$^{-1}$ (2--10\,keV, unabsorbed), corresponding to a luminosity of $\sim 4\times10^{37}$ erg\,s$^{-1}$ at 13 kpc, which is typical of SFXT outbursts. 

\begin{minipage}{.95\textwidth}
  \begin{minipage}[b]{0.46\textwidth}
    \centering
\includegraphics[angle=0,height=0.7\textwidth]{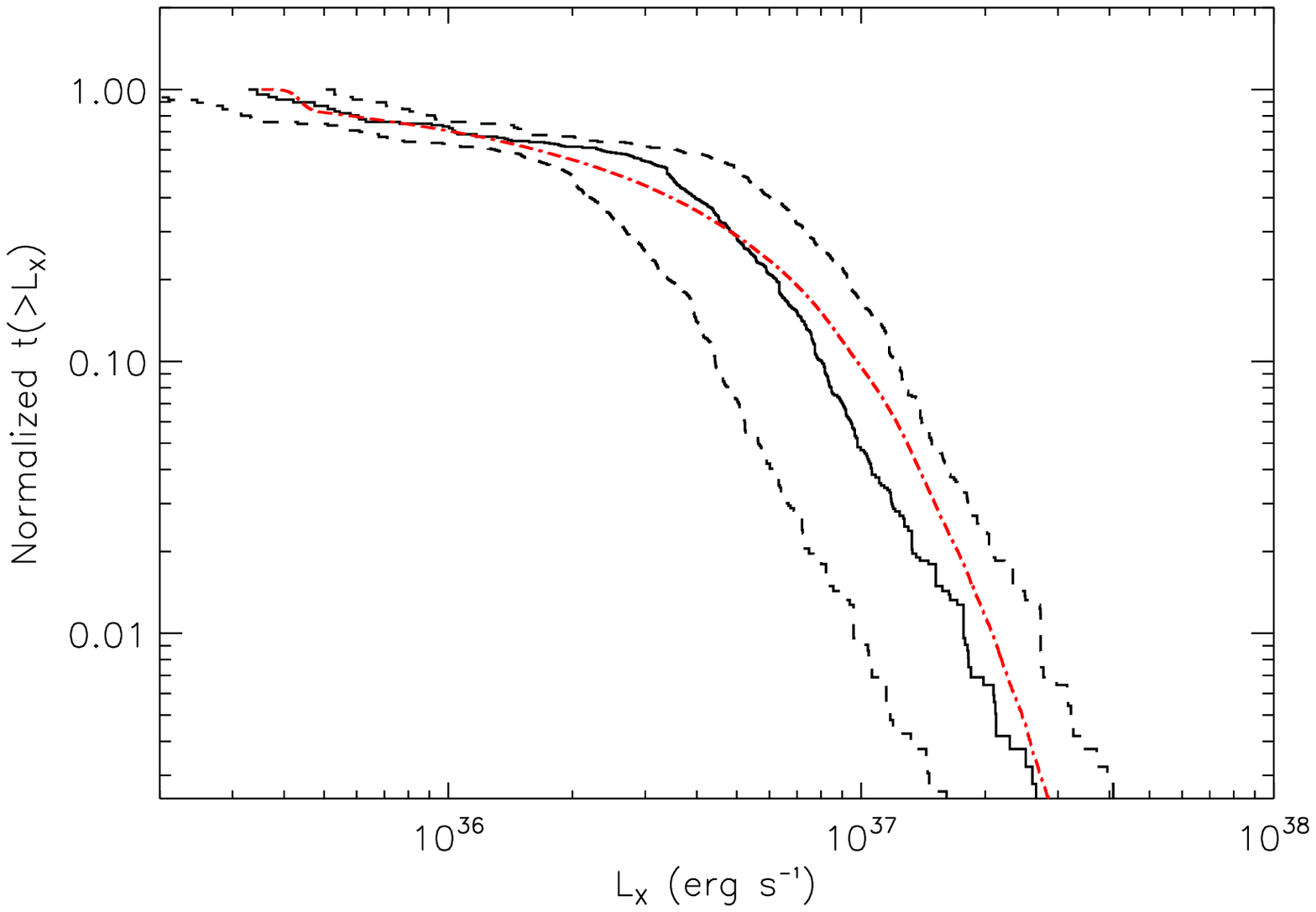}
    \captionof{table}{\textbf{FIGURE~3.}~Observed and computed cumulative luminosity distribution; see text.
}
  \end{minipage}
  \hfill
  \begin{minipage}[b]{0.53\textwidth}
    \centering
    \begin{footnotesize}
    \begin{tabular}{ccc}
\hline
Parameter  & min. value & max. value \\
\hline
$\dot{M}_{\rm tot}$ & $4 \times 10^{-7}$\,M$_\odot$\,yr$^{-1}$ & $8 \times 10^{-7}$\,M$_\odot$\,yr$^{-1}$ \\
$f$        & $0.85$  & $0.95$ \\
$v_\infty$  & $800$\,km\,s$^{-1}$ & $1300$\,km\,s$^{-1}$ \\
$\beta$    & $0.8$ & $1.2$ \\
$\zeta$    &  $0.8$ & $1.6$ \\
$\gamma$   & $-1.5$ & $-0.5$ \\
$M_{\rm a}$ & $5 \times 10^{16}$\,g & $5 \times 10^{19}$\,g \\
$M_{\rm b}$ & $5 \times 10^{19}$\,g & $10^{21}$\,g \\
\hline
      \end{tabular}
      \end{footnotesize}
      \captionof{table}{\textbf{TABLE~1.}~Wind parameter values for \src. \label{igr16418:tab:wind-parameters}}
    \end{minipage}
  \end{minipage}
\vspace{0.5cm}
%
%
%
%

To reproduce the observed X-ray variability, we applied a clumpy wind model 
\citep{Ducci2009}, which we briefly summarize here. 
This model assumes that the wind of OB supergiants
is inhomogeneous, composed of dense clumps surrounded by a hotter
and homogeneous wind. Clumps follow a power-law mass distribution,
$p(M_{\mathrm{cl}})=k (M_{\mathrm{cl}}/M_{\mathrm{a}})^{-\zeta}$,
where $k=(1 - \zeta)\,M_{\rm a}^{-\zeta}/(M_{\rm b}^{1-\zeta}-M_{\rm a}^{1-\zeta})$ 
is the normalization constant,
$M_{\rm cl}$ is the mass of a clump,
and $M_{\rm a}$ and $M_{\rm b}$ define the extrema of the mass range.
Assuming a spherical geometry for the clumps,
the initial clump dimension distribution is given by 
$\dot{N}_{M_{cl}} \propto R_{\rm cl}^{\gamma}$\,clumps\,s$^{-1}$,
where $R_{\rm cl}$ is the radius of the clump.
The clump mass loss rate $\dot{M}_{\rm cl}$ is related to the total
mass loss rate $\dot{M}_{\rm tot}$ by means of 
$f=\dot{M}_{\rm cl} / \dot{M}_{\rm tot}$,
the fraction of wind mass in the form of clumps.
Clumps and the homogeneous wind follow the $\beta$\emph{-velocity} law
\begin{math} \label{legge_velocita}
v(r) = v_{\infty}(1 - 0.9983{R_{\rm OB}}/{r})^{\beta}
\end{math}
\citep{Castor1975},
where $\beta$ is a constant in the range $\sim$$0.5-1.5$,
$R_{\rm OB}$ is the radius of the supergiant and 
$v_{\infty}$ is the terminal wind velocity.

%
%
Each \sw\ snapshot 
has a duration of $\sim 1$\,ks, for a total of up to $\sim 5$\,ks per day,
which is the typical timescale for the 
flare durations observed in SFXTs.
For this reason, we cannot establish for \src\ the number of observed flares and their duration,
therefore we cannot use this information
to compare the observed lightcurve with that calculated with
the clumpy wind model. 
However, it is possible to compare the observed cumulative 
luminosity distribution out of eclipse
with that calculated with the model.
For the former, 
the \emph{Swift}/XRT count rate, measured in the $0.3$--$10$\,keV,
was converted to the $0.1$--$100$\,keV luminosity using the 
average spectral parameters obtained by \citep{Walter2006,Ducci2010}, and for a distance of 13\,kpc.
For the latter, we adopted the physical parameters for the O8.5I 
primary and the neutron star secondary reported above, and reasonable
the force multiplier parameters for a O8.5I star \citep[see][]{romano12}: $k=0.375$, $\alpha=0.522$, $\delta=0.099$. 

Fig.\,3 shows the XRT cumulative luminosity distribution (solid line) with its 90\,\% uncertainties (dashed lines) and also the cumulative luminosity distribution (red dot-dashed line) calculated with the clumpy wind model by adopting the following parameter values: mass loss rate $\dot{M}_{\rm tot}=6 \times 10^{-7}$\,M$_\odot$\,yr$^{-1}$, terminal velocity $v_\infty=1000$\,km\,s$^{-1}$, $\beta=1$, fraction of mass lost in clumps $f=0.9$, mass distribution power-law index $\zeta = 1.1$,  power-law index of the initial clump dimension distribution $\gamma =-1$, minimum clump mass $M_{\rm a}=5 \times 10^{18}$\,g and maximum clump mass $M_{\rm b}=10^{20}$\,g. Further acceptable solutions can be found with the wind parameters in the ranges reported in Table\,1. 

The wind parameters obtained for \src\ are roughly 
in agreement with those typical of HMXBs,
with the exception of $f$ and $v_\infty$.
The obtained low value of the terminal velocity
can be explained as follows: when the wind is highly ionized by the X-ray photons
emitted by the neutron star,
the high ionization modifies the dynamics of the line-driven stellar wind 
of OB supergiants (especially in close binary systems),
leading to a reduction of the wind velocity in the direction of the neutron star
\citep[e.g.][]{Stevens1990}.
The mass loss rate is lower than the typical
mass loss rate of O8.5I stars, which is of the order of 
$\approx 2 \times 10^{-6}$\,M$_\odot$\,yr$^{-1}$ \citep{Vink2000},
but in agreement with the hypothesis that 
the mass loss rates derived from homogeneous-wind model
measurements with the H$\alpha$ method are overestimated
by a factor of 2--10 if the wind is clumpy \citep[e.g.][]{Lepine2008}.
However, we cannot exclude the possibility that the X-ray variability
of \src\ is not totally due to the accretion
of an inhomogeneous wind. In fact, other mechanisms could be at work,
like e.g.\ transient accretion discs and 
intermittent accretion flow onto a neutron star.

\begin{theacknowledgments}
See the Swift Supergiant Fast X-ray Transients Project page at \url{http://www.ifc.inaf.it/sfxt/}. We received financial contribution from ASI-INAF grants I/009/10/0 and I/004/11/0.
\end{theacknowledgments}
\vspace{-0.2cm}





\IfFileExists{\jobname.bbl}{}
 {\typeout{}
  \typeout{******************************************}
  \typeout{** Please run "bibtex \jobname" to optain}
  \typeout{** the bibliography and then re-run LaTeX}
  \typeout{** twice to fix the references!}
  \typeout{******************************************}
  \typeout{}
 }

\end{document}